\documentclass[reprint,aps,prb,floatfix,twocolumn,superscriptaddress,showpacs,amsmath,amssymb,showpacs,]{revtex4-1}
\usepackage{graphicx}%
\usepackage{bm}%
\usepackage{amsmath, nccmath}
\usepackage[USenglish]{babel}
\usepackage{xcolor} 
\usepackage[utf8]{inputenc}
\newcommand{\Kappa}{\mbox{\Large $\kappa$}}

\begin{document}

\author{Viktor Christiansson}
\affiliation{Department of Physics, University of Fribourg, 1700 Fribourg, Switzerland}
\author{Francesco Petocchi} 
\affiliation{Department of Physics, University of Fribourg, 1700 Fribourg, Switzerland}
\author{Philipp Werner }
\affiliation{Department of Physics, University of Fribourg, 1700 Fribourg, Switzerland}
\date{\today}
\title{Superconductivity in black phosphorus and the role of dynamical screening}

\begin{abstract}
Simple cubic phosphorus exhibits superconductivity with a maximum $T_c$ of up to 12 K under pressure. The pressure dependence of $T_c$ cannot be consistently explained with a simple electron-phonon mechanism, which has stimulated investigations into the role of electronic correlations and plasmonic contributions. Here, we solve the gap equation of density functional theory for superconductors using different electron-electron and electron-phonon contributions to the kernel. We find that the phonon contribution alone yields an overestimation of $T_c$, while the addition of the static 
electronic contribution results in an underestimation. Taking into account the full frequency dependence of the screened interaction, the one-shot $GW$ approximation predicts $T_c$ values in good agreement with the experiments in the pressure range appropriate for the cubic phase. 
We also explore the use of quasi-particle bands in the calculation of the electronic and phononic kernels, and show that this modification significantly improves $T_c$ in the high-pressure region.
\end{abstract}

\maketitle

\section{\label{sec:Introduction}Introduction}

\begin{figure}[ht!] 
\begin{centering}
\includegraphics[width=\columnwidth]{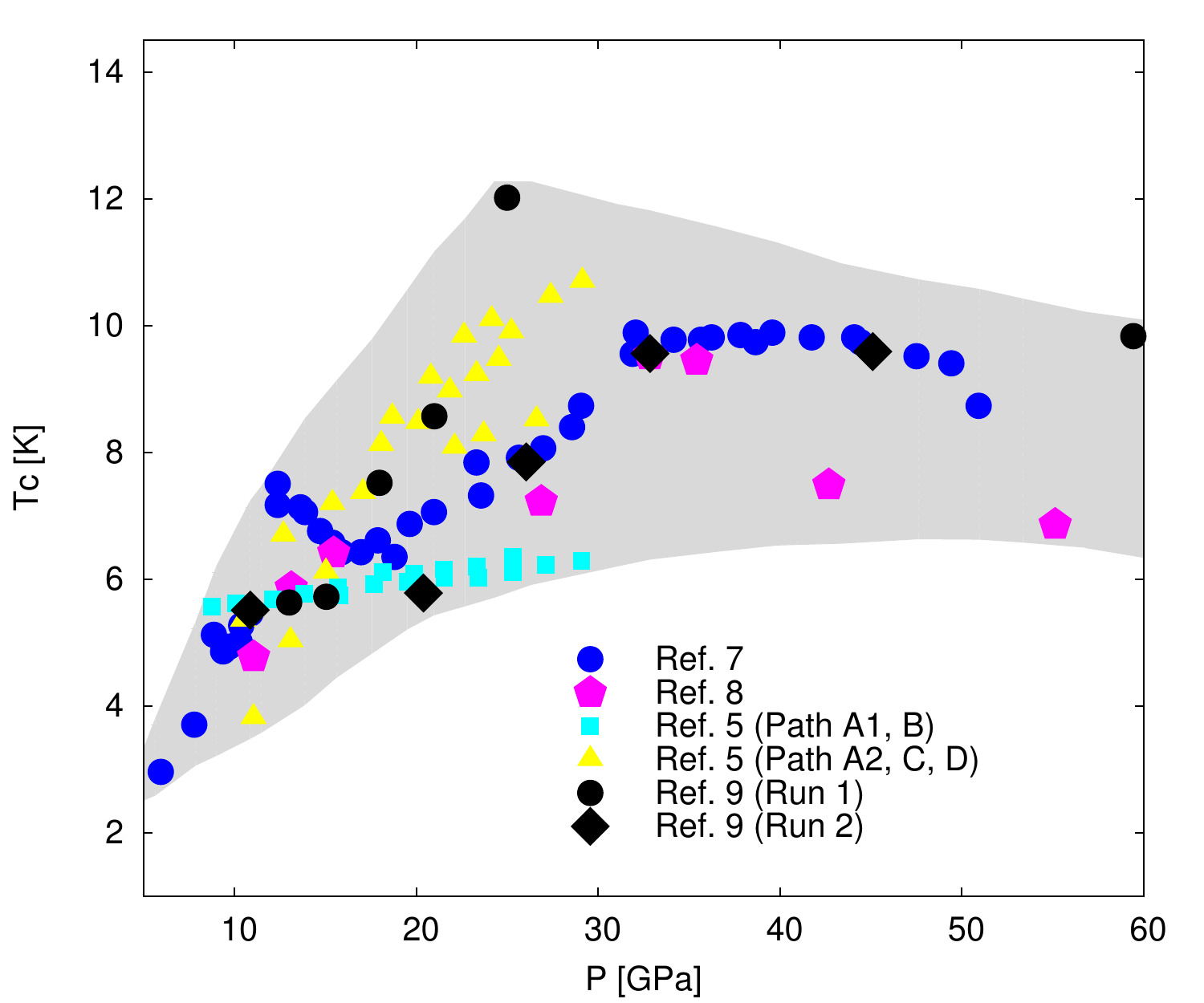}
\par\end{centering}
\caption{Pressure dependence of the experimental superconducting critical temperature $T_c$. The data have been extracted from Refs.~\onlinecite{Guo2017,Karuzawa2002,Shirotani1988,Livas2017}, as indicated in the legend. The shaded area outlines the spread of measured values of $T_c$, and will be used to test our theoretical results.  \label{Fig:ExpTc}
}
\end{figure}

Black phosphorus at ambient conditions is a layered semiconductor with a narrow gap. It turns into a metallic simple cubic phase at a pressure of about 10 GPa,\cite{Kikegawa1983} and the cubic structure has been reported to remain stable up to 107 GPa.\cite{Akahama1999} At low temperatures, superconductivity is observed for pressures above 5 GPa, and the pressure dependence of the superconducting critical temperature $T_c$ has been the subject of numerous experimental and theoretical studies. Despite this effort, a thorough theoretical understanding of the pairing mechanism and  superconductivity in cubic phase phosphorus is still lacking.  
On the experimental side, the situation is further complicated by the wide variation in the measured $T_c$ values, depending on the experimental protocol, as indicated in Fig.~\ref{Fig:ExpTc}.
For example, it was shown by Kawamura {\it et al.}\cite{Kawamura1984,Kawamura1985,Shirotani1988} that the precise pressure-temperature path has significant effects on the pressure dependence of the superconducting critical temperature. With a certain choice of thermodynamical path, they obtained an almost constant $T_c$ with increasing pressure, whereas another path produced a more rapidly increasing $T_c$. Later experiments by Wittig {\it et al.}\cite{Wittig1985} showed a valley-like structure at lower pressures, which agrees with a similar finding by Guo {\it et al.}  in Ref.~\onlinecite{Guo2017}. The latter results, however, predicted a roughly constant $T_c$ at higher pressures, forming a ridge-like structure, whereas the former found a decreasing $T_c$ after a maximum near 23 GPa.
Yet another form of the $T_c$ versus pressure curve was reported by Karuzawa {\it et al.},\cite{Karuzawa2002} who measured a pressure dependence with a single maximum of the $T_c$ around 32 GPa.

Different mechanisms have been proposed to explain the remarkable robustness of $T_c$ under high pressure and various features in the experimental measurements. 
However, no consistent theory for the full pressure dependence and for the pairing mechanism has so far emerged. 
Based on measurements of the Hall coefficient, Guo {\it et al.} \cite{Guo2017} interpreted the valley structure in the $T_c$-versus-pressure diagram around 17 GPa as originating from a Lifshitz transition in the simple cubic phase. 
This has later been supported by the theoretical analysis of Wu {\it et al.},\cite{Wu2018} who performed density functional theory (DFT)\cite{Hohenberg1964,Kohn1965} calculations of the Fermi surface, reporting 
two subsequent Lifshitz transitions occurring in the pressure range where, 
using the McMillan equation,\cite{McMillan1968,Allen1975} a $T_c$ valley was also predicted.
Flores-Livas {\it et al.},\cite{Livas2017} using an {\it ab-inito} calculation based on density functional theory for superconductors (SCDFT),\cite{Oliveira1988,Luders2005,Marques2005} and the static interaction coming from a random-phase approximation (RPA) \cite{Pines1963} calculation, also argued that the rapid increase of $T_c$ was due to a Lifshitz transition. 
These authors furthermore argued that the Lifshitz transition did not occur within the simple cubic phase, but rather as a result of a structural transition from the rhombohedral to the simple cubic phase, while Ref.~\onlinecite{Guo2017} reports a transition to the simple-cubic phase around 10-13 GPa. 
Their analysis further suggested that the discrepancies seen in the experimental results up to 25 GPa can be explained by the co-existence of different structural phases.
Neither of the theoretical studies was, however, able to reproduce the plateau found experimentally by Guo {\it et al.} for pressures between $30$ and $50$ GPa, as well as Flores-Livas {\it et al.} in their second experimental run. Since the static interaction used in the earlier work by Flores-Livas {\it et al.} predicted a peak instead of a ridge, Wu {\it et al.} hypothesized that plasmonic contributions from the inclusion of the full frequency dependent interaction \cite{Akashi2013} may provide an additional effective attraction which stabilizes the $T_c$ at higher pressures.
 
The goal of our study is to go beyond the previous predictions based on the McMillan formula or static RPA interactions by considering also the dynamic (frequency-dependent) contribution to the screened interaction in SCDFT, as proposed by Akashi {\it et al.} in Ref.~\onlinecite{Akashi2013}. Using this {\it ab-initio} scheme, we will  study different levels of approximations to the electron-electron interaction, including RPA, one-shot and self-consistent $GW$,\cite{Hedin1965} and $GW$ plus extended dynamical mean-field theory ($GW$+EDMFT).\cite{Biermann2003,Ayral2013,Boehnke2016,Nilsson2017} For better consistency between the calculation of the interaction and the SCDFT scheme used to predict $T_c$, we furthermore explore a quasi-particle extension of the formalism. In contrast to the previous SCDFT study,\cite{Livas2017} we assume the simple cubic phase in the whole pressure range, since this is the experimentally observed structure for pressures in the most interesting region of the possible valley-ridge $T_c$  structure. 

The paper is organized as follows. In Sec.~\ref{sec:Method} we detail the SCDFT formalism and the methods we use to obtain the screened interaction. In Sec.~\ref{sec:Results} we present our results and compare them to the available experimental data and earlier theoretical studies, while in Sec.~\ref{sec:Summary} we summarize our conclusions.

\section{\label{sec:Method}Method}

\subsection{General remarks}

In this section, we introduce the methods we use for predicting $T_c$ within the framework of SCDFT, as well as some computational details. First, the bandstructure of cubic phase phosphorus, obtained from a DFT calculation will be presented in Sec.~\ref{sec:DFT}, where we also show the theoretical pressure-volume curve which is used to compare to experiments. In Sec.~\ref{sec:many-body} we introduce our seven-band model and the methods used to compute the dynamically screened interactions for it, while the estimation of the phononic contribution is discussed in Sec.~\ref{sec:phonons}. In Sec.~\ref{sec:SCDFT} we explain the calculation of $T_c$ by the SCDFT formalism.

\begin{figure}[t]
\begin{centering}
\includegraphics[width=0.95\columnwidth]{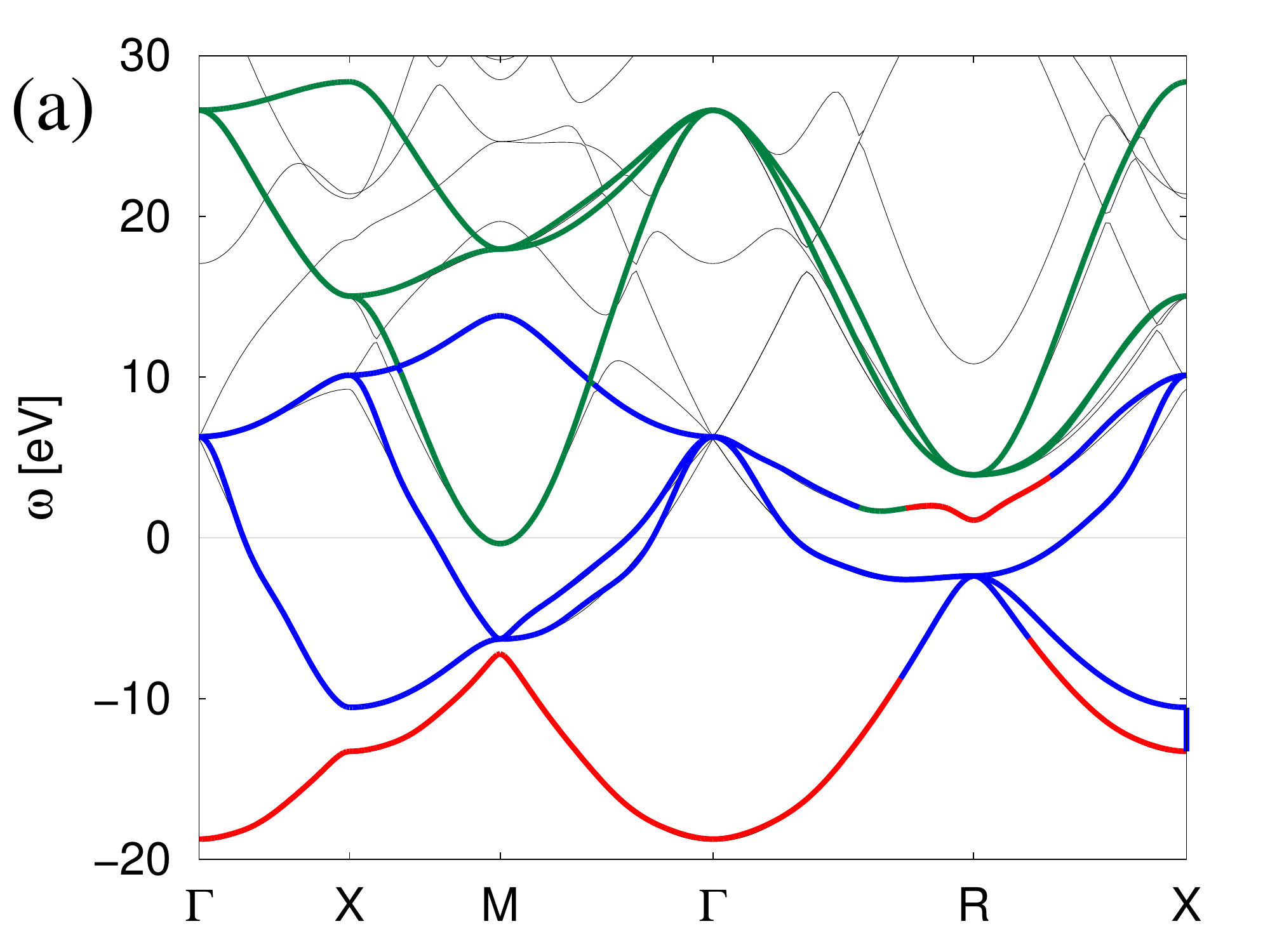}
\includegraphics[width=0.95\columnwidth]{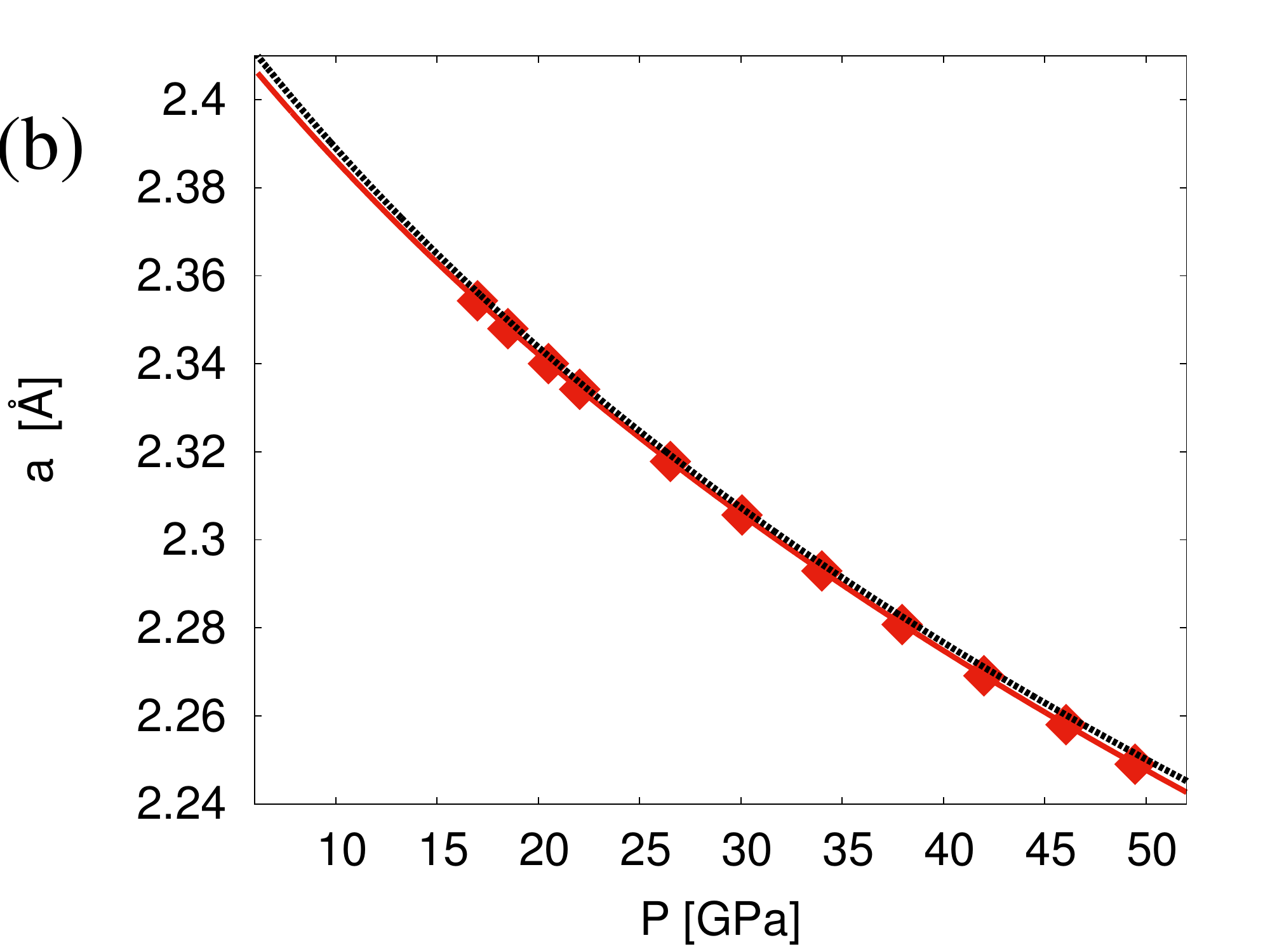}
\par\end{centering}
\caption{
(a) DFT band structure along high-symmetry lines for simple cubic phosphorus at a pressure of $\sim 38$ GPa. The colored superimposed lines define the low-energy subspace spanned by seven bands: red, blue, and green indicate majority $s$-, $p$-, and $t_{2g}$-like orbital characters, respectively. 
(b) Theoretical pressure dependence of the lattice constant $a$ used for the comparison between experimental and theoretical data from FLEUR (red) and ELK (dotted black). The diamonds mark the pressures used for the $T_c$ estimates. 
\label{Fig:BS}}
\end{figure}

\subsection{Band structure for cubic phase phosphorus}\label{sec:DFT}

All the calculations start with a DFT calculation of the electronic structure of phosphorus in the simple cubic phase. The generalized-gradient approximation (GGA),\cite{Perdew1996} as implemented in the full-potential linearized augmented plane-wave code FLEUR,\cite{Fleurcode} is used to obtain the {\it ab-initio} band structure on a $32\times 32 \times 32$ $\bf k$-point grid. The band structure along high-symmetry lines for one pressure ($P\approx 38$ GPa) is shown in Fig.~\ref{Fig:BS}(a). In agreement with earlier theoretical works,\cite{Aoki1987,Rajagopalan1989,Chan2013,Wu2018} upon increasing pressure we observe two consecutive Lifshitz transitions; the first is associated with the disappearance of a pocket of predominantly $s$ character around the R momentum at 21-22 GPa, which is followed by a second one at 22-23 GPa with a pocket of $d$ character appearing around the M momentum. The orbital characters of the bands near the Fermi energy are superimposed in Fig.~\ref{Fig:BS}(a), showing the dominant contribution to the bands crossing the Fermi energy to come from orbitals of $p$ character. 

For the comparison with the experimental results, we also fit the energy-volume data of the DFT calculations using the Vinet equation of state \cite{Vinet1986} 
to obtain the theoretical pressure corresponding to a lattice constant in the simple cubic phase, see Fig.~\ref{Fig:BS}(b).

\subsection{Dynamically screened interactions}\label{sec:many-body}

To move beyond DFT, we define a low-energy model consisting of seven orbitals using Maximally localized Wannier functions \cite{Marzari1997} from the Wannier90 library,\cite{Mostofi2008} starting from orbitals of $3s,\,p,$ and $t_{2g}$ character. This provides a low-energy model in good agreement with the DFT band structure in a large energy range around the Fermi energy, as shown in  Fig.~\ref{Fig:BS}(a). The band character is indicated by the majority contribution of the $s$-, $p$-, and $t_{2g}$-like Wannier orbitals to the model bands. 

We subsequently perform a systematic downfolding of the full band structure to the low-energy space by means of a constrained random-phase approximation (cRPA) \cite{Aryasetiawan2004} and a one-shot $GW$ calculation.\cite{Hedin1965} 
In the $GW$ approximation,\cite{Hedin1965} the self-energy is expanded to first order in the screened Coulomb interaction $W$, which produces a set of coupled equations for the Green's function $G$, self-energy $\Sigma$, screened interaction $W$ and polarization $\Pi$. 

In reciprocal space and at zero temperature the self-energy is given by
\begin{equation}\label{Eq:SigmaGW}
\Sigma_{\bf k}(\omega)=\frac{i}{2\pi}\sum_{\bf q} \int \textrm{d}\omega' G_{{\bf k}-{\bf q}}(\omega+\omega')W_{{\bf q}}(\omega').
\end{equation}
The screened interaction $W_{{\bf q}}$ is calculated by screening the bare Coulomb interaction $v_{{\bf q}}$ by the polarization function $\Pi_{{\bf q}}$,
\begin{equation}\label{Eq:WGW}
W_{{\bf q}}(\omega)=v_{{\bf q}} + v_{{\bf q}} \Pi_{{\bf q}}(\omega) W_{{\bf q}}(\omega),
\end{equation}
where $\Pi_{{\bf q}}$ is calculated within RPA as 
\begin{equation}\label{Eq:PiGW}
\Pi_{{\bf q}}(\omega) = -\frac{i}{2\pi} \sum_{{\bf k}} \int \textrm{d}\omega' G_{{\bf k}}(\omega') G_{{\bf k}-{\bf q}}(\omega'-\omega).
\end{equation}

The $GW$ approximation requires as initial input a non-interacting Green's function, $G^0_{\bf k}$, which is commonly taken from a DFT calculation. This replaces initially $G_{\bf k}$ in Eqs.~(\ref{Eq:SigmaGW})-(\ref{Eq:PiGW}), which yield the Green's function of the one-shot $GW$ (or $G^0W^0$) approximation,
\begin{equation}\label{Eq:GGW}
G_{\bf k}^{-1}(\omega)= (G^0_{\bf k})^{-1}(\omega) - \Sigma_{\bf k}(\omega).
\end{equation}
Starting from the DFT derived $G_{\bf k}^0$, Eqs.~(\ref{Eq:SigmaGW})-(\ref{Eq:GGW}) can be iterated by using the updated Green's function $G$ in the next iteration, and if this is repeated until self-consistency, the method is referred to as self-consistent $GW$ (sc$GW$). In practice, however, good or even better results are obtained by one-shot $GW$ in many cases, unless the self-consistency loop is modified.\cite{Schilfgaarde2006}

Through a $G^0W^0$ calculation in the full space, using the disentangled band structure, \footnote{To disentangle the 7 bands in the low-energy model, we used an outer window up to 40 eV and an inner window between -5:5 eV in the Wannierization.} we obtain the embedding self-energy for the seven bands of our model, $\Sigma^\text{embedding}_{{\bf k}}(\omega)$, which, together with $G^0_{{\bf k}}$, yields the effective bare propagators in the model subspace.\cite{Boehnke2016,Nilsson2017}
Similarly, the cRPA method is used to calculate the effective bare interaction within the model space: the bands inside the low-energy subspace, in our case the 7 band model, are excluded from the polarization in the $G^0W^0$ calculation, Eq.~(\ref{Eq:PiGW}). A similar equation to Eq.~(\ref{Eq:WGW}) is then obtained for the partially screened interaction
\begin{equation}
U_{{\bf q}}(\omega)=v_{{\bf q}} + v_{{\bf q}} \Pi^{r}_{{\bf q}}(\omega) U_{{\bf q}}(\omega),
\end{equation}
where the superscript $r$ indicates that the summations in the formula for the RPA polarization should be done over all bands except for transitions within the model subspace. The resulting frequency dependent interaction $U_{{\bf q}}(\omega)$ represents the effective bare interaction for the model space. By subsequently screening $U_{{\bf q}}$ with the polarization $\Pi^{\textrm{model}}_{{\bf q}}$ from the previously excluded bands that define the model subspace, the fully screened interaction $W_{{\bf q}}$ in Eq.~\eqref{Eq:WGW} is recovered. 

The $G^0W^0$ and cRPA calculations were performed with the SPEX code \cite{Friedrich2010} at zero temperature. A $8\times8\times8$ {\bf k}-grid was used and DFT bands up to 100 eV were included in the calculation for both the polarization function and the self-energy.  Having obtained the  effective bare propagators $G^0_{{\bf k}}$ and interactions $U_{{\bf q}}$ in the model space, we employed several approximate methods to compute the screened interaction needed for the SCDFT formalism (see Sec.~\ref{sec:SCDFT}). 
These will be briefly explained in the following. 

For the RPA and $GW$ variants of the screened interaction, we have evaluated $W_{{\bf q}}$ in Eq.~\eqref{Eq:WGW} using the RPA-type polarization function (Eq.~\eqref{Eq:PiGW}) with the following choices of $G^0_{{\bf k}}$ and interaction parameters:
\begin{enumerate}
\item The DFT non-interacting Green's function, rotated from the Kohn-Sham basis to the Wannier basis, and the bare interaction $v_{\bf q}$. This defines $W^{\textrm{RPA}}$.
\item \label{Enum:G0W0} The bare effective propagator in the model space obtained with Eq.~\eqref{Eq:GGW} using the DFT $G^0_{{\bf k}}$ and $\Sigma^\text{embedding}_{{\bf k}}$ from the $G^0W^0$ calculation. Furthermore, $v_{\bf q}$ is replaced by the partially screened interaction $U_{\bf q}(\omega)$. This procedure defines $W^{G^0W^0}$.
\item \label{Enum:scGW} Similar to point \ref{Enum:G0W0}, but with the Green's function obtained in a self-consistent manner from the finite-temperature equivalents of Eqs.~(\ref{Eq:SigmaGW})-(\ref{Eq:GGW}) (see e.~g. Ref.~\onlinecite{Nilsson2017}) within the model subspace. Again, $U_{\bf q}(i\omega_n)$ replaces $v_{\bf q}$ as the bare interaction in Eq.~\eqref{Eq:WGW}. This calculation defines $W^{\textrm{sc}GW}$. 
\end{enumerate}
The full expression for the Green's function in method \ref{Enum:scGW} is
\begin{align}
G^{-1}_{{\bf k}}=& i\omega_n + \mu -\varepsilon_{{\bf k}}^{\textrm{DFT}} + V^\text{XC}_{{\bf k}} \nonumber \\
	& - \left(\Sigma_{{\bf k}}^{G^0W^0} - \Sigma_{{\bf k}}^{G^0W^0}\big|_\textrm{model} + \Sigma_{{\bf k}}^{\textrm{sc}GW}\big|_\textrm{model} \right).
\end{align}
The $G^0W^0$ calculation provides the contribution from the states outside the model, with the exchange-correlation potential $V^\text{XC}_{{\bf k}}$ from the DFT calculation and the self-energy from a one-shot $G^0W^0$ calculation within the model space ($\Sigma_{{\bf k}}^{G^0W^0}|_\textrm{model}$) removed to avoid double countings of interaction contributions.\cite{Boehnke2016,Nilsson2017}

In addition we have calculated the screened interaction obtained from a fully self-consistent $GW$+EDMFT simulation \cite{Biermann2003,Boehnke2016,Nilsson2017} ($W^{GW\textrm{+EDMFT}}$) where we update all seven orbitals with the local vertex corrections from EDMFT,\cite{Metzner1989,Georges1996,Sun2002} and in a multitier $GW$+EDMFT simulation, where we limit the EDMFT corrections to only the $t_{2g}$-like orbitals ($W^{\textrm{multitier}}$).\cite{Nilsson2017} 

We perform the $GW$+EDMFT and sc$GW$ calculations 
at nonzero temperatures, by first analytically continuing the initial zero-temperature calculations to the Matsubara axis. Due to the large computational cost associated with these non-local, frequency dependent calculations, we are here limited to a relatively high temperature of $T\approx 380$ K (inverse temperature $\beta=30 $ eV$^{-1}$). We have checked that the screened interactions obtained in this way do not display significant changes when temperature is further lowered, so that we use them as an approximation also for the low temperature system. 

\begin{figure}[ht] 
\begin{centering}
\includegraphics[width=0.49\textwidth]{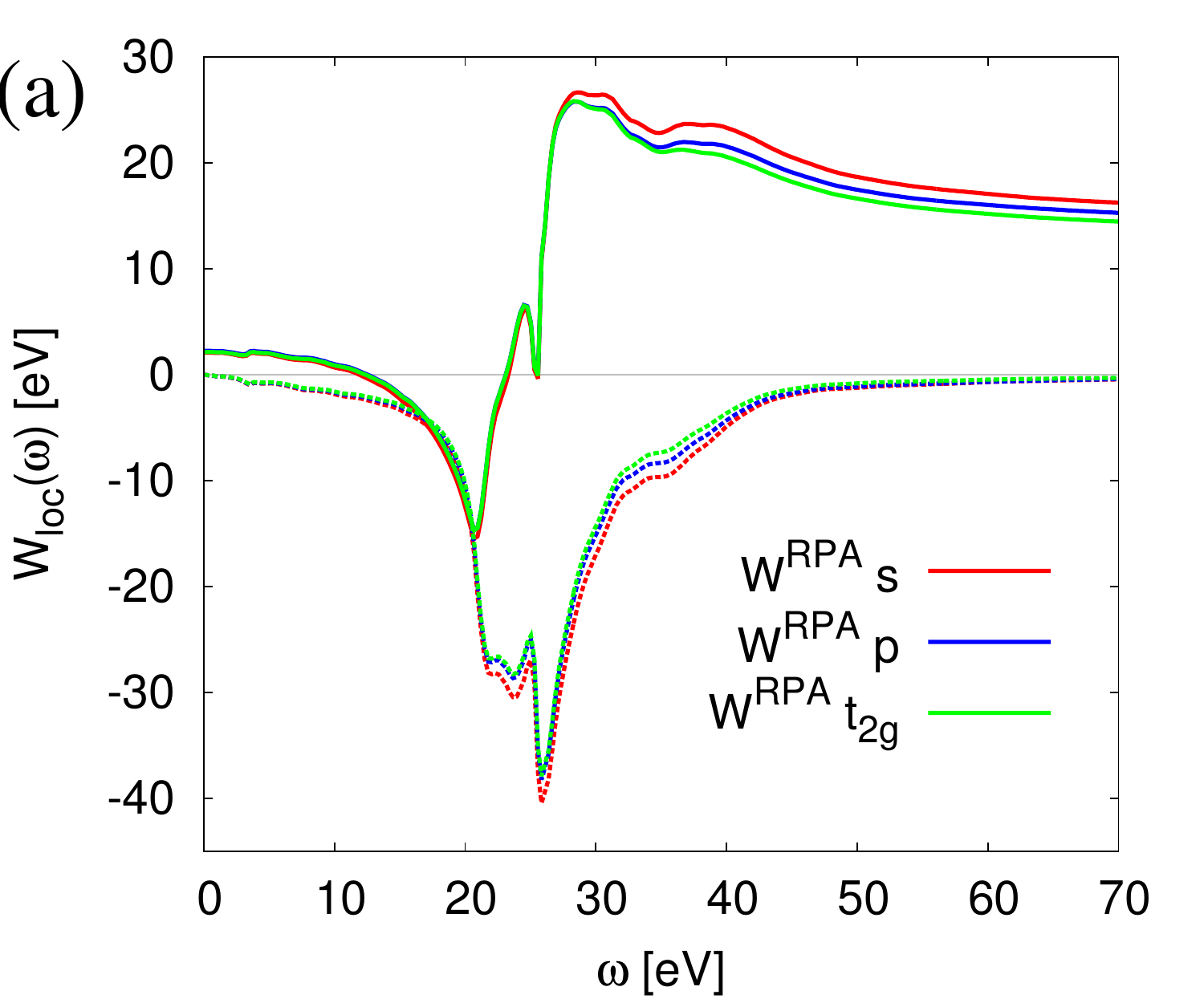}
\includegraphics[width=0.49\textwidth]{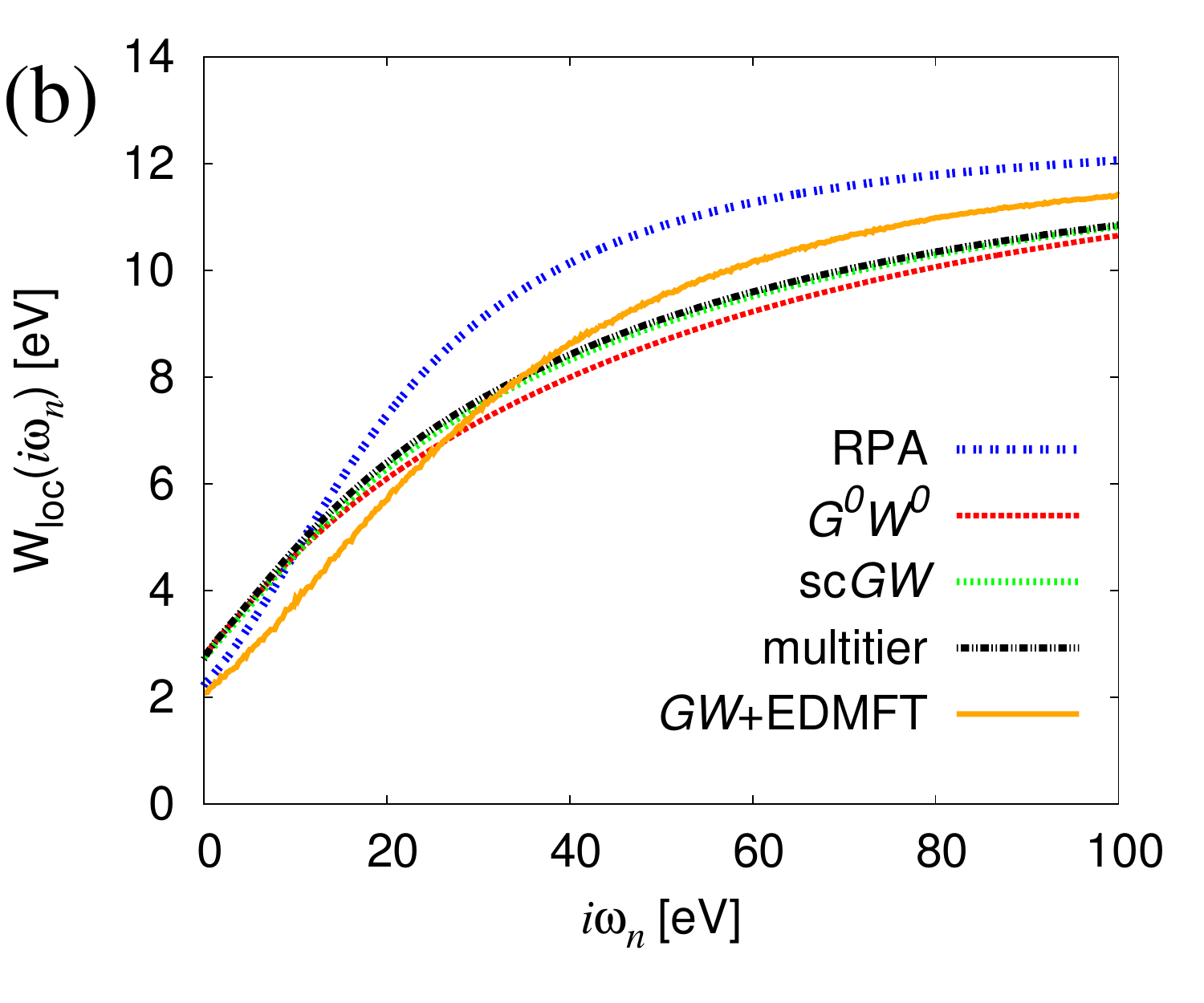}
\par\end{centering}
\caption{Panel (a): Frequency dependence of the fully screened local interaction, $W_\text{loc}(\omega)$, calculated within RPA on the real axis. The full and dashed lines show the real and imaginary parts for the orbitals with $s$- (red), $p$- (blue), and t$_{2g}$-character (green).
Panel (b): Comparison of the local fully screened interaction on the Matsubara axis for the orbital with t$_{2g}$-character, calculated within RPA (blue), $G^0W^0$ (red), sc$GW$ (green), multitier $GW$+EDMFT (black), and $GW$+EDMFT (orange). 
\label{Fig:screenedW}
}
\end{figure}

In Fig.~\ref{Fig:screenedW}(b) we show the $t_{2g}$-like component of the interaction from the different methods. The SCDFT formalism requires the interaction on the Matsubara axis, but for comparison we also show in panel (a) the real-frequency dependence of the local $W^{\textrm{RPA}}$ which we can obtain directly from the $G^0W^0$ downfolding without analytical continuation. 
This figure displays the RPA interactions for the $s$-, $p$-, and $t_{2g}$-like components, which are very similar.
For the calculation of $T_c$ in Sec.~\ref{sec:SCDFT} we retain all the off-diagonal components of $W$.
The method employed in this work does not allow us to directly identify the relative importance of the contributions from the different bands.

\subsection{Eliashberg function}\label{sec:phonons}

The phononic contribution to the superconductivity enters the SCDFT formalism used in this work via the Eliashberg function,\cite{Allen1972} which is calculated as 
\begin{align}\label{Eq:a2F}
\alpha^2F(\omega) =& \frac{1}{N(0)}\sum_{\lambda,{\bf q}}\sum_{nn',{\bf k}}\left| g_{\lambda,{\bf q}}^{n{\bf k},n'{\bf (k+q)}} \right|^2\nonumber\\
&\times\delta(\varepsilon_{n{\bf{k}}})\delta(\varepsilon_{n' {\bf k}+{\bf q}})\delta(\omega-\omega_{\lambda{\bf q}}).
\end{align}
Here $\varepsilon_{n{\bf{k}}}$ denotes the one-particle energies (measured from the Fermi energy) for the states $(n,{\bf k})$ obtained from the DFT band structure, the phonon frequencies are $\omega_{\lambda {\bf q}}$ for wave vector ${\bf q}$ and mode $\lambda$, and $N(0)$ is the density of states at the Fermi energy. 
The matrix elements of the electron-phonon coupling constants are given by 
\begin{equation}
g_{\lambda,{\bf q}}^{n{\bf k},n'{\bf (k+q)}} = \frac{1}{\sqrt{2M\omega_{\lambda{\bf q}}}} \left \langle n',({\bf k}+{\bf q}) \right| \delta_{{\bf q}}^\lambda \, V^\text{KS} \left| n,{\bf k} \right\rangle,
\end{equation}
where the variation of the Kohn-Sham potential with respect to the displacements is denoted by $\delta_{{\bf q}}^\lambda \, V^\text{KS}$, and $M$ is the mass of the atom.

The phononic contributions were calculated with the supercell method as implemented in the ELK code \cite{elkcode} using the GGA functional on a $8\times8\times8$ $\bf q$-grid and a $32\times32\times32$ $\bf k$-grid. The interpolated fine grids used for the $\bf k$- and $\bf q$-integrations in Eq.~\eqref{Eq:a2F} were taken to be $128\times128\times128$ and $200\times200\times200$ respectively, which was sufficient for good convergence. Since the volume-pressure curve is in good agreement between FLEUR and ELK, as shown in Fig.~\ref{Fig:BS}(b), and as both are full-potential all-electron codes, the electronic and phononic contributions to the SCDFT calculations described in the next section should be compatible.

\subsection{\label{sec:SCDFT}SCDFT}
\begin{figure}[ht]
\includegraphics[width=\columnwidth]{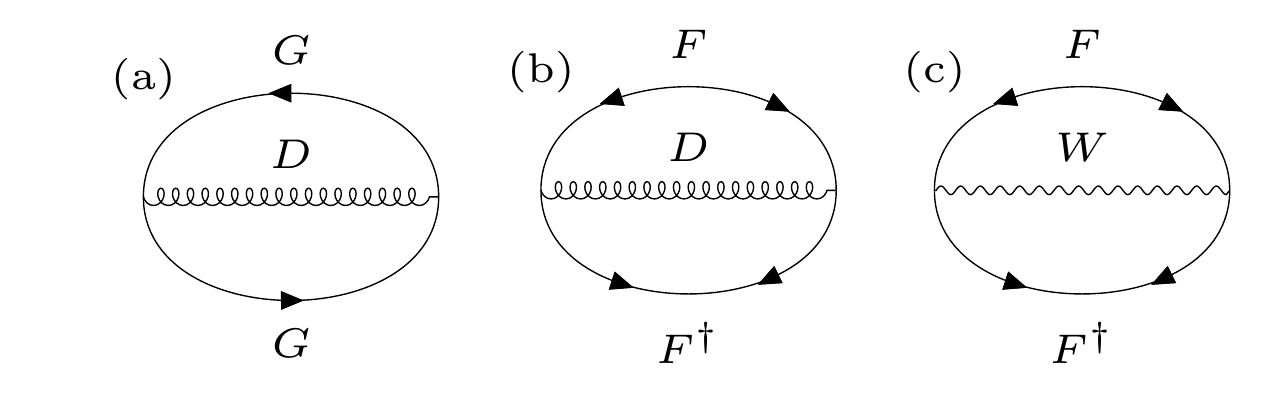}
\caption{\label{Fig:Feynman} Diagrammatic representation of the exchange-correlation kernels used in this work: (a) and (b) show the electron-phonon contributions to $\mathcal{Z}^{\textrm{ph}}$ and $\Kappa^{\textrm{ph}}$, respectively, while (c) illustrates the electron-electron contribution to $\Kappa^{\textrm{el}}$ (the kernels are defined in the text). The curled line represents the phonon propagator $D$, the wavy line the screened Coulomb interaction $W$, single-arrowed lines the electronic Green's function $G$, and double-arrowed lines the anomalous Green's functions $F$ and $F^\dagger$. 
}
\end{figure}

Density functional theory for superconductors\cite{Oliveira1988,Luders2005,Marques2005} is a formalism which allows to predict the superconducting critical temperature $T_c$ from first principles. 
The $T_c$ is estimated from the vanishing of the gap function $\Delta_{n{\bf{k}}}$, which is obtained as the self-consistent solution of the gap equation 
\begin{equation}
\Delta_{n\bf{k}}=-\mathcal{Z}_{n\bf{k}}\Delta_{n\bf{k}}-\frac{1}{2}\sum_{n'\bf{k}'}\Kappa_{n{\bf{k}},n'\bf{k'}}\frac{\tanh[(\beta/2)E_{n'\bf{k}'}]}{E_{n'\bf{k}'}}\Delta_{n'\bf{k}'}.
\end{equation}
This equation involves the exchange-correlation kernels $\mathcal{Z}$ and $\Kappa$, the energies 
$E_{n{\bf{k}}}=\sqrt{\varepsilon_{n{\bf{k}}}^2+\Delta_{n{\bf{k}}}^2}$, and the inverse temperature $\beta$. 
In this study, the diagonal $\mathcal{Z}$ term is assumed to consist of only the electron-phonon contribution $\mathcal{Z}=\mathcal{Z}^{\textrm{ph}}$, whereas for the $\Kappa$ kernel we include both the electron-phonon and electron-electron contributions: $\Kappa=\Kappa^{\textrm{ph}}+\Kappa^{\textrm{el}}$. A diagrammatic representation of the exchange-correlation functionals is shown in Fig.~\ref{Fig:Feynman}.

Within the $\bf k$-dependent formalism derived in Ref.~\onlinecite{Luders2005}, the expressions for the electron-phonon kernels  are
\begin{align}
\mathcal{Z}^{\textrm{ph}}_{n{\bf k}} =& \frac{1}{\tanh[(\beta/2)\varepsilon_{n{\bf k}}]}\sum_{n {\bf k'}}\sum_{\lambda {\bf q}}\left| g^{n{\bf k},n'{\bf k'}}_{\lambda{\bf q}} \right|^2\nonumber\\ 
&\times\left[ J(\varepsilon_{n{\bf k}},\varepsilon_{n'{\bf k'}},\omega_{\lambda{\bf q}}) + J(\varepsilon_{n{\bf k}},-\varepsilon_{n'{\bf k'}},\omega_{\lambda{\bf q}}) \right]
\end{align}
and
\begin{align}
\Kappa^{\textrm{ph}}_{n{\bf k}n'{\bf k'}} =& \frac{1}{\tanh[(\beta/2)\varepsilon_{n{\bf k}}]}\frac{1}{\tanh[(\beta/2)\varepsilon_{n'{\bf k'}}]}\sum_{\lambda {\bf q}}\left| g^{n{\bf k},n'{\bf k'}}_{\lambda{\bf q}} \right|^2\nonumber\\
& \times \left[ I(\varepsilon_{n{\bf k}},\varepsilon_{n'{\bf k'}},\omega_{\lambda{\bf q}}) - I(\varepsilon_{n{\bf k}},-\varepsilon_{n'{\bf k'}},\omega_{\lambda{\bf q}}) \right],
\end{align}
with the functions $I$ and $J$ defined in terms of the Fermi-Dirac ($n_F(\varepsilon)$) and Bose-Einstein ($n_B(\omega)$) distributions as

\begin{align}
&I(\varepsilon,\varepsilon',\omega)= n_F(\varepsilon)n_F(\varepsilon')n_B(\omega)\nonumber\\
&\quad\times \left( \frac{e^{\beta \varepsilon}-e^{\beta(\varepsilon'+\omega)}}{\varepsilon-\varepsilon'-\omega} - \frac{e^{\beta \varepsilon'}-e^{\beta(\varepsilon+\omega)}}{\varepsilon-\varepsilon'+\omega} \right),\\
%
&J(\varepsilon,\varepsilon',\omega)= \tilde{J}(\varepsilon,\varepsilon',\omega)-\tilde{J}(\varepsilon,\varepsilon',-\omega),\\
&\tilde{J}(\varepsilon,\varepsilon',\omega)=-\frac{n_F(\varepsilon)+n_B(\omega)}{\varepsilon-\varepsilon'-\omega}\nonumber\\
&\quad\times \left( \frac{n_F(\varepsilon')-n_F(\varepsilon-\omega)}{\varepsilon-\varepsilon'-\omega} - \beta n_F(\varepsilon-\omega)n_F(-\varepsilon+\omega) \right).
\end{align}

For the electronic contribution to the $\Kappa$ kernel, we employ the fully  frequency dependent interaction kernel proposed in Ref.~\onlinecite{Akashi2013}, which has been shown in previous studies to properly account for the electron-electron interaction effects. This term is separated into the static and dynamic contributions, $\Kappa^{\mathrm{el}}=\Kappa^{\mathrm{el},\,\mathrm{static}}+\Kappa^{\mathrm{el},\,\mathrm{dynamic}}$, to identify their respective effects.
The static part of the kernel is given by the static screened interaction
\begin{equation}
\Kappa_{n{\bf{k}},n'{\bf{k}'}}^{\text{el},\,\mathrm{static}}=W_{n{\bf{k}},n'{\bf{k}'}}(0),
\end{equation}
and the dynamic (frequency dependent) part by
\begin{align}
&\Kappa_{n{\bf{k}},n'{\bf{k}'}}^{\mathrm{el},\,\mathrm{dyn}}=\lim_{\Delta_{n{\bf k}} \to 0} \frac{1}{\tanh(E_{n{\bf k}}\beta/2)}\frac{1}{\tanh(E_{n'{\bf k}'}\beta/2)}\nonumber\\
&\quad\times\frac{1}{\beta^2}\sum_{\omega_1\omega_2}F_{n{\bf k}}(i\omega_1)F_{n'{\bf k}'}(i\omega_2) \left[ W^{\mathrm{dyn}}_{n{\bf k}n'{\bf k}'}(i\omega_1-i\omega_2)   \right].
\end{align}
The anomalous (electronic) Green's function $F_{n{\bf k}}$ is given by 
\begin{equation}
F_{n{\bf k}}(i\omega_j)=\frac{1}{i\omega_j + E_{n{\bf k}}}-\frac{1}{i\omega_j - E_{n'{\bf k}'}},
\end{equation}
where the $\omega_j$ are fermionic Matsubara frequencies, and for simplicity we have introduced the notation $W^{\mathrm{dyn}}_{n{\bf k}n'{\bf k}'}(i\omega_1-i\omega_2) = W_{n{\bf k}n'{\bf k}'}(i\omega_1-i\omega_2) - W_{n{\bf k}n'{\bf k}'}(0)$.
This can be simplified,\cite{Tsutsumi2020} using the variable transformation $i\nu=i(\omega_1-\omega_2)$, to the following expression which only requires a summation over a single bosonic frequency $\nu$, 
\begin{widetext}
\begin{align}
\Kappa_{n{\bf{k}},n'{\bf{k}'}}^{\text{el},\,\mathrm{dyn}}=&\lim_{\Delta_{n {\bf k} \to 0}}\frac{1}{\tanh(\beta/2 E_{n{\bf{k}}})}\frac{1}{\tanh(\beta/2 E_{n'{\bf{k}}'})}\frac{1}{\beta}\sum_\nu\left[W^\text{dyn}_{n{\bf{k}},n'{\bf{k}'}}(i\nu)\right] \nonumber\\ 
&\times\left( \frac{2(E_{n{\bf{k}}}-E_{n'{\bf{k}}'})}{(E_{n{\bf{k}}}-E_{n'{\bf{k}}'})^2+\nu^2}(n_F(E_{n{\bf{k}}})-n_F(E_{n'{\bf{k}}'}))+\frac{2(E_{n{\bf{k}}}+E_{n'{\bf{k}}'})}{(E_{n{\bf{k}}}+E_{n'{\bf{k}}'})^2+\nu^2}(n_F(-E_{n{\bf{k}}})-n_F(E_{n'{\bf{k}}'})) \right).
\end{align}
\end{widetext}
The screened interaction $W$ has commonly been computed using RPA or the adiabatic local density approximation,\cite{Zangwill1980,Gross1985} whereas in the present study we will compare the results for the screened interactions obtained by the different diagrammatic schemes described in Sec.~\ref{sec:many-body}.

The phononic contribution requires a sufficiently dense ${\bf k}$-grid close to the Fermi energy for convergence in the low-energy regime. Often this is handled by a random sampling method, with a higher density of $\bf k$-points close to the Fermi energy to ensure a sufficient resolution.\cite{Marques2005,Akashi2012} Due to the prohibitively large computational cost for sc$GW$ and $GW$+EDMFT, which scales quadratically with the number of $\bf k$-points, we are unable to calculate $W_{n{\bf k}n{\bf k'}}$ directly in this way and instead would have to interpolate from a coarse grid. For this reason, we resorted to the energy-averaged formalism,\cite{Marques2005} where the electron-electron interaction is first evaluated using the {\it ab-initio} methods on a coarse grid, as described in section \ref{sec:many-body}. We thereafter obtain the kernels by averaging over iso-energetic surfaces, using analytical expressions for the phononic parts while the electronic kernel has to be obtained numerically. 

The energy-averaged version of the gap equation takes the form\cite{Marques2005}
\begin{align}
&\Delta(\varepsilon)=-\mathcal{Z}(\varepsilon)\Delta(\varepsilon)\nonumber\\
& \quad - \frac{1}{2}\int_{-\mu}^\infty d\varepsilon' N(\varepsilon')\Kappa(\varepsilon,\varepsilon')\frac{\tanh[(\beta/2) E']}{E'} \Delta(\varepsilon')
\label{Eq:Gapeq}
\end{align}
and the energy-averaged phononic kernel $\mathcal{Z}^\text{ph}$ becomes
\begin{align}
&\mathcal{Z}^\text{ph}(\varepsilon)=-\frac{1}{\tanh(\beta/2\varepsilon)}\int_{-\mu}^\infty d \varepsilon' \nonumber\\
& \quad\times\int d\omega \alpha^2F(\omega) \left[ J(\varepsilon,\varepsilon',\omega)+J(\varepsilon,-\varepsilon',\omega)\right],
\end{align}
where $\mu$ is the chemical potential, and the Eliashberg function, $\alpha F(\omega)$, has been defined in Sec.~\ref{sec:phonons}.
As previously noted, the $\Kappa$ kernel consists of two parts in our calculations: the electron-phonon ($\Kappa^\text{ph}$) and the electron-electron ($\Kappa^\text{el}$) contributions. The electron-phonon kernel within the energy-averaged formalism is expressed as
\begin{align}
&\Kappa^\text{ph}(\varepsilon,\varepsilon')=\frac{2}{\tanh(\varepsilon\beta/2)\tanh(\varepsilon'\beta/2)}\frac{1}{N(0)} \nonumber\\
&\quad \times \int d\omega \alpha^2F(\omega) \left[ I(\varepsilon,\varepsilon',\omega)-I(\varepsilon,-\varepsilon',\omega)\right],
\end{align}
with $N(\varepsilon)$ the density of states, whereas the integrals over the iso-energetic surfaces must be done numerically for the electron-electron contribution:
\begin{equation}
\Kappa^{\text{el}}(\varepsilon,\varepsilon')=\frac{1}{N(\varepsilon)N(\varepsilon')}\sum_{n{\bf{k}},n'{\bf{k}'}}\delta(\varepsilon-\varepsilon_{n{\bf k}})\delta(\varepsilon'-\varepsilon_{n {\bf k'}})\Kappa^{\text{el}}_{n{\bf k},n'{\bf k}}.
\end{equation}
The final expressions used in this work are
\begin{widetext}
\begin{equation}
\Kappa^{\text{el},\,\mathrm{static}}(\varepsilon,\varepsilon')=\frac{1}{N(\varepsilon)N(\varepsilon')}\sum_{n{\bf{k}},n'{\bf{k}'}}\delta(\varepsilon-\varepsilon_{n{\bf k}})\delta(\varepsilon'-\varepsilon_{n {\bf k'}})W_{n{\bf{k}},n'{\bf{k}'}}(0),
\label{Eq:Kelstatic}
\end{equation}
\begin{equation}
\begin{split}
\Kappa^{\text{el},\,\mathrm{dyn}}(\varepsilon,\varepsilon')&=\frac{1}{\tanh(\beta/2 \varepsilon)}\frac{1}{\tanh(\beta/2  \varepsilon')}\frac{1}{\beta}\sum_\nu\left[\frac{1}{N(\varepsilon)N(\varepsilon')}\sum_{n{\bf{k}},n'{\bf{k}'}}\delta(\varepsilon-\varepsilon_{n{\bf k}})\delta(\varepsilon'-\varepsilon_{n {\bf k'}}) W^\text{dyn}_{n{\bf{k}},n'{\bf{k}'}}(i\nu) \right] \\ 
&\times\left( \frac{2( \varepsilon- \varepsilon')}{( \varepsilon- \varepsilon')^2+\nu^2}(n_F( \varepsilon)-n_F( \varepsilon'))+\frac{2( \varepsilon+ \varepsilon')}{( \varepsilon+ \varepsilon')^2+\nu^2}(n_F(- \varepsilon)-n_F( \varepsilon')) \right),
\end{split}
\label{Eq:Keldynamic}
\end{equation}
\end{widetext}
where we made use of the tetrahedron method\cite{Jepson1971,Lehmann1972} to carry out the $\bf k$-integrations numerically, to obtain $\left[W^\text{dyn}(i\nu)\right]$. 

The methods used to compute $W$ at nonzero temperatures are limited to high temperatures (described in Sec.~\ref{sec:many-body}) compared to the observed $T_c$. We assume that the same $W$ can be used for all $T_c$ estimates.
Noting that $W$ is an even function in $\nu$,  to perform the frequency summation, we replace\cite{Akashi2015}
$
\sum_\nu \rightarrow \frac{\beta}{\pi}\big( \int_0^{\nu_\text{max}} d\nu + \int_{\nu_\text{max}}^{\infty} \big),
$
and introduce a frequency cutoff $\nu_\text{max}=300$ eV, after which the tail of the interaction is assumed to be constant. This can be justified by checking that the high-energy behavior has approximately reached the bare value. The second integral can then be evaluated analytically and contributes 
\begin{equation*}
\begin{split}
&\left[W^\text{dyn}(\nu^{\text{max}})\right] \left( 1-\frac{2}{\pi}\arctan\left[\frac{\nu_\text{max}}{(\varepsilon\pm\varepsilon')}\right]\right)\left( n_F(\mp \varepsilon) - n_F(\varepsilon') \right) 
\end{split}
\end{equation*}
to the kernel. The remaining integral up to the cutoff is treated numerically using the change of variables $\nu=(\varepsilon\pm\varepsilon')(1+y)/(1-y)$.\cite{Kawamura2017}

\section{\label{sec:Results}Results}

\subsection{General remarks}

We computed the critical temperatures from the vanishing of the superconducting gap in Eq.~\eqref{Eq:Gapeq} at pressures ranging from $\sim$17 to 50 GPa. For pressures below 17 GPa we obtained significant imaginary phonon frequencies, indicative of a structural instability, in agreement with previous DFT calculations for simple cubic phosphorus in this pressure range.\cite{Chan2013,Wu2018} These instabilities essentially disappeared around $20.5$ GPa in our calculations. To investigate effects related to the method used to obtain the interaction $W$ entering the electron-electron contribution, we evaluated the pressure dependence of $T_c$ using the methods listed in Sec.~\ref{sec:many-body} for both the static and fully-dynamic $\Kappa^\text{el}$ kernels. The results will be presented in order of increasing complexity in the kernels considered, starting from a purely phononic kernel, then describing the effects of further including the static electronic part,  followed by the simulations with the fully dynamic kernel. Finally we explore a strategy that allows to include the correlation effects not only through the interaction, but also through the phononic contribution, 
by replacing the non-interacting bandstructure used in  SCDFT with the quasi-particle bandstructure.

\subsection{Phonon contribution only}

\begin{figure}[t] 
\begin{centering}
\includegraphics[width=0.48\textwidth]{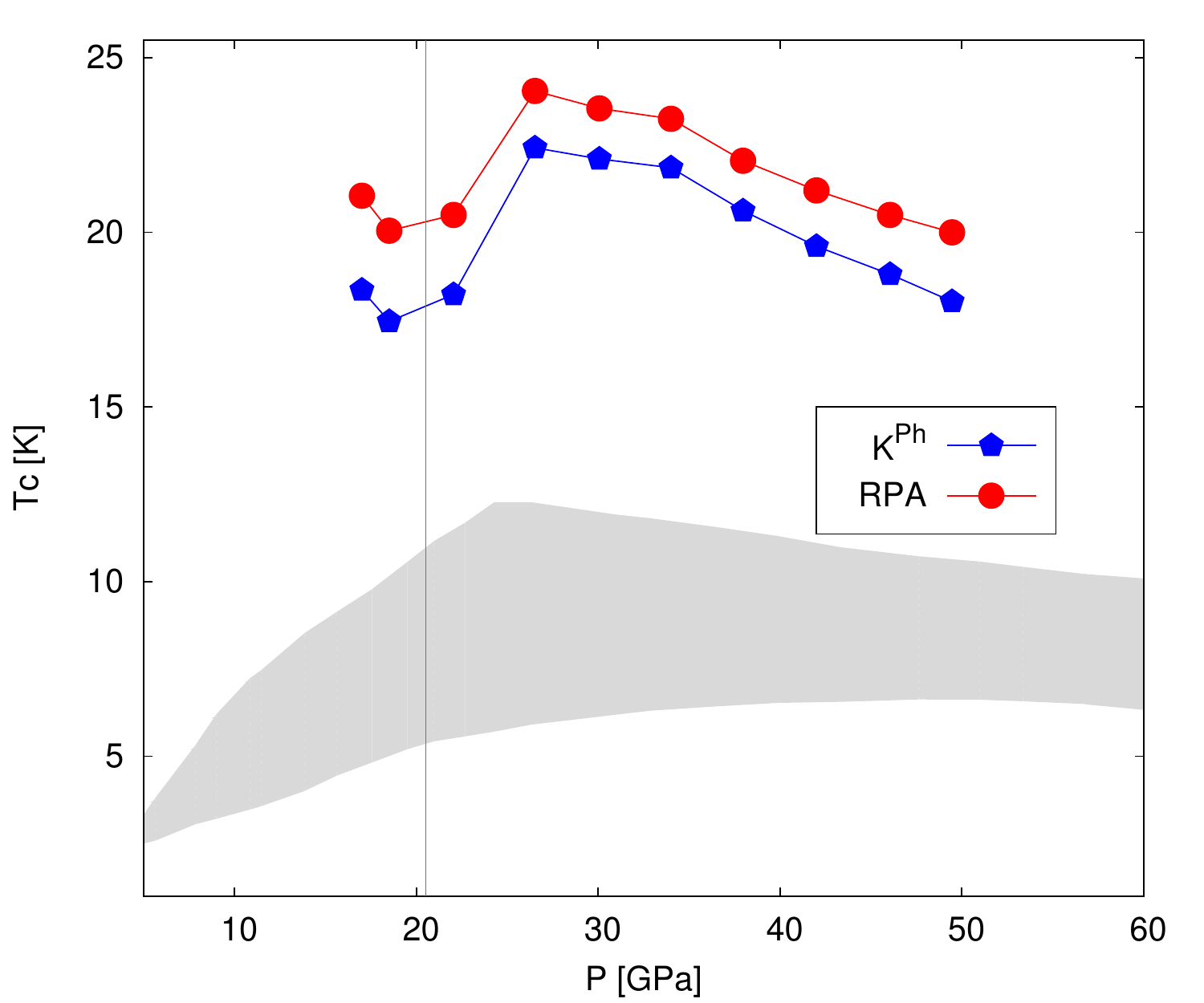}
\par\end{centering}
\caption{Theoretically calculated critical temperature $T_c$ as a function of pressure compared to the experimental results (gray shading). To the left of the  vertical line we observe a structural instability in our phonon calculations. The blue line shows the result with the phonon kernels only and the red line the result with the (additional) electronic contribution from RPA. 
Here, we consider the full frequency dependence of the electronic contribution.  
\label{Fig:TcTheory_phonon}
}
\end{figure}

The pressure dependence of the $T_c$ values obtained with only the phonon contributions are shown by the blue line in Fig.~\ref{Fig:TcTheory_phonon}. The phonon-only approximation severely overestimates $T_c$ by more than a factor of two compared to the experimental values. This is because such a calculation misses cancellation effects between the phononic and electronic contributions, which are known to suppress $T_c$.\cite{Marques2005} 
In more phenomenological theories this problem is usually addressed by introducing effective parameters, for example in the McMillan equation through the effective interaction parameter $\mu^*$.\cite{McMillan1968,Allen1975} Within the SCDFT formalism employed here, a similar effect is produced by adding the contribution from the static electronic kernel, $\Kappa^{\text{el},\,\mathrm{static}}$, which drastically reduces the predicted $T_c$. This underestimation in turn is to different extents mitigated by including the contributions from the dynamical kernel, as will be discussed in the following. 

\subsection{Phonon contribution plus static electronic contribution}

\begin{figure}[t]
\begin{centering}
\includegraphics[width=0.48\textwidth]{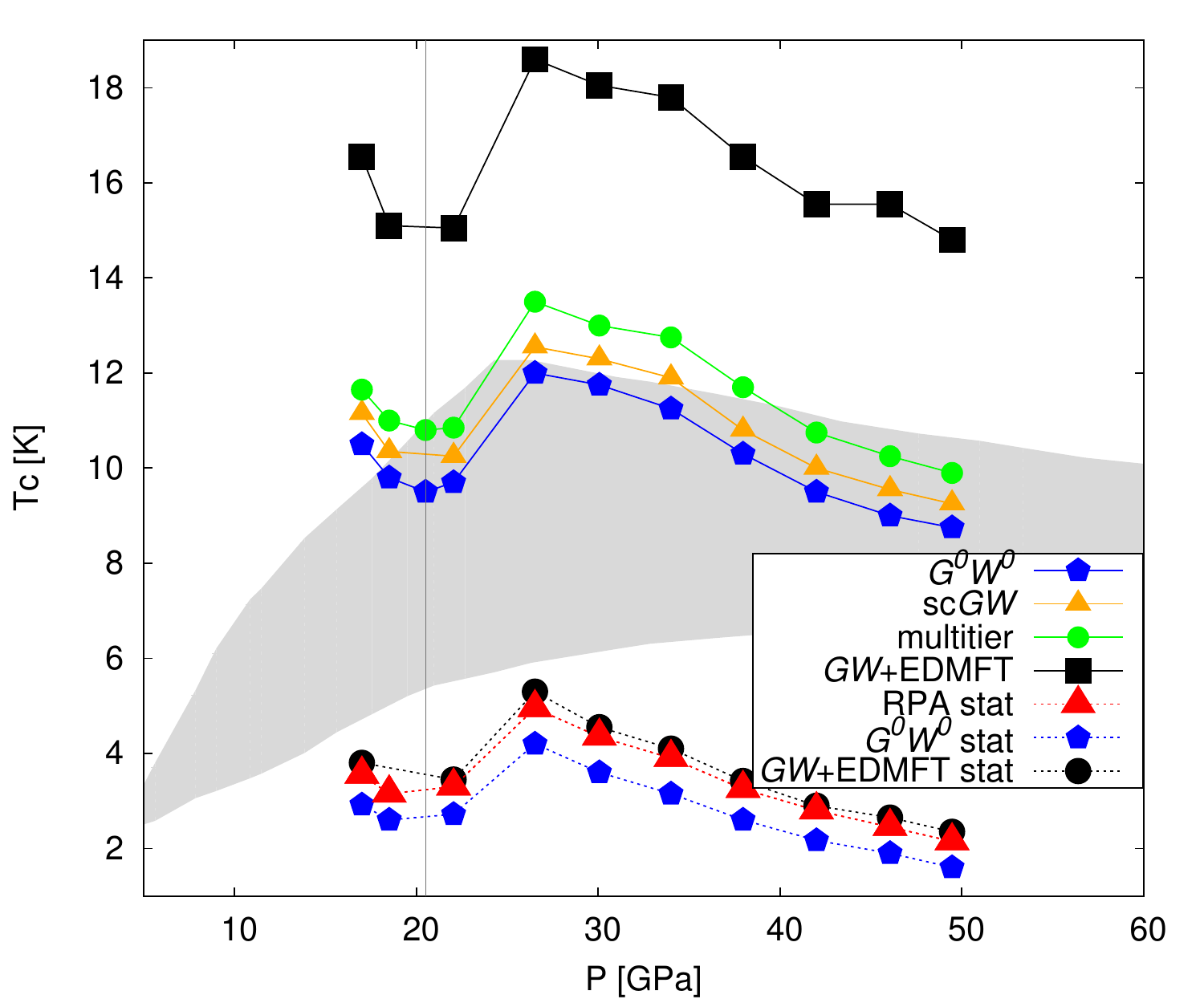}
\par\end{centering}
\caption{Theoretically calculated critical temperature $T_c$ as a function of pressure $P$ compared to the experimental results (gray shading). To the left of the vertical line we observe a structural instability in our phonon calculations. Dashed lines show results obtained with the phononic plus static electronic kernels, and the full lines the results which additionally include the dynamical contribution, i.e., which consider the full frequency dependence of the electronic contributions.
\label{Fig:TcTheory_static}}
\end{figure}

Considering only the static interaction in the calculation of the electronic kernel, $\Kappa^{\text{el}}=\Kappa^{\text{el},\,\mathrm{static}}$, we obtain a severe underestimation of the critical temperature for all methods employed in this work, as shown in Fig.~\ref{Fig:TcTheory_static} by the dashed lines. The static results from one-shot $GW$, sc$GW$, and multitier $GW$+EDMFT are in very close agreement and only one representative pressure-$T_c$ curve is shown for these (labelled $G^0W^0$ static). 

The predicted $T_c$ from $GW$+EDMFT and RPA is slightly higher, but still severely underestimated compared to the experimental measurements. This is a first indication that adding the local EDMFT corrections to the $s$- and $p$-like orbitals produces significant differences, compared to the other $GW$-based schemes, a result which will be discussed further in the following sections. Within the current approach, the different results can be understood from the behavior of the static value of the local screened interaction presented in Fig.~\ref{Fig:screenedW}: the low-frequency screening turns out to be more pronounced for the $GW$+EDMFT and RPA methods, leading to a larger reduction of the static electron-electron interaction. Since $\Kappa^{\text{el},\,\mathrm{static}}$ is a positive and approximately constant quantity, it partly cancels the $T_c$ enhancing contributions from the oppositely signed $\Kappa^{\text{ph}}$ close to the Fermi energy, which in turn produces the observed differences in the suppression of $T_c$.

Let us compare these results to phenomenological theories predicting a $T_c$ valley in this region.\cite{Wu2018} When only considering the static electronic kernel in our scheme, the static value of the non-local, orbital-dependent interaction plays the role of an effective parameter controlling the critical temperature and its pressure dependence. Instead of using it as an adjustable parameter, however, it is calculated here in a fully {\it ab-initio} way using a range of methods. All the used methods predict the formation of a valley structure around $20$ GPa in the simple cubic phase, a few GPa above the value where such a structure has been observed in experiments,\cite{Wittig1985,Guo2017} although at severely underestimated $T_c$ values. Flores-Livas {\it et al.}\cite{Livas2017} also predicted a small valley in this region, although their calculations were in the A7 phase instead of the simple cubic phase in this pressure range.

The fact that our calculated values of $T_c$ consistently underestimate the experimental values irrespective of the method demonstrates the importance of including the dynamic kernel, and hence the retardation effect from single-particle and collective charge excitations (plasmons),\cite{Akashi2013} for an accurate description of the critical temperature.

\subsection{Phonon contribution plus static and dynamic electronic contributions}

Within our formalism, the inclusion of the fully dynamical electronic contribution enhances $T_c$ and brings the calculated results into the range of the experimentally measured values, see the symbols connected by full lines in Fig.~\ref{Fig:TcTheory_static}.
In addition, compared to the static case, the $T_c$ curves differ more between the methods, with an almost rigid shift between the results for one-shot $GW$, sc$GW$, and multitier $GW$+EDMFT. This demonstrates the sensitivity of the SCDFT approach on the dynamic contribution to the electronic kernel, and allows us to identify the most suitable method for the present system. The best agreement with experiment is found for one-shot $GW$. This is consistent with other studies on weakly correlated systems, where fully self-consistent $GW$ is found to perform worse than one-shot $GW$.\cite{Holm1998}

As discussed in the previous section, the static value of $W$ is almost the same for one-shot $GW$, sc$GW$ and multitier $GW$+EDMFT. We can relate the observed differences to the frequency dependence of the local screened interaction shown in Fig.~\ref{Fig:screenedW}. The dynamical screening varies more widely between the methods, with the significantly reduced RPA screening at higher energies (compared to one-shot $GW$) together with the over-screened static interaction being responsible for the overestimation of the $T_c$ (see red line in Fig.~\ref{Fig:TcTheory_phonon} and note the different scale compared to Fig.~\ref{Fig:TcTheory_static}). Also the $GW$+EDMFT interaction shows two frequency regimes; at low frequencies it over-estimates the screening, compared to one-shot $GW$, while at high energies it underestimates it. The net result is again a less accurate $T_c$ (black line in Fig.~\ref{Fig:TcTheory_static}).
It should also be noted that, unlike the static kernel, the dynamic one is no longer approximately constant as a function of energy due to the additional factors in Eq.~\eqref{Eq:Keldynamic}. This prohibits us from drawing any conclusions based on the local interaction only.

The fact that $GW$+EDMFT worsens the agreement with experiments indicates that the local self-energy and polarization contributions are overestimated in $GW$+EDMFT, relative to the nonlocal ones. As was discussed in Ref.~\onlinecite{Nilsson2017},
for weakly correlated materials with strong nonlocal screening, corrections beyond the RPA-type diagrams would be needed for the nonlocal part. The replacement of the local polarization by the EDMFT result, but the restriction of the nonlocal polarization to a simple bubble, produces a mismatch between local and nonlocal screening effects, and an incorrect estimation of the interaction. 
In the case of one-shot $GW$, the local and nonlocal polarizations are treated on equal footing, and the estimated $T_c$ is improved accordingly. 

The $T_c$ is substantially different when only the self-energies and polarizations of the $t_{2g}$-orbitals are corrected with the local quantities from EDFMT within the multitier $GW$+EDMFT formalism, which produces results which are more similar to one-shot $GW$ and sc$GW$. This indicates two things: firstly that local corrections to the (almost empty) $d$-orbitals are of minor importance in this material for the calculation of $W$ and the description of the pressure dependence of the critical temperature, and secondly that the effects of treating the $s$ and $p$ states with the local EDMFT corrections is primarily responsible for the overestimation of $T_c$ in $GW$+EDMFT. This agrees with the previous discussion on the importance of not adding the full local contributions to states whose screening is not well described by a bubble approximation to the nonlocal diagrams.

In contrast to the very sharp $T_c$ valley found by Wu {\it et al.},\cite{Wu2018} where the maxima on the two sides roughly coincide, we observe a more shallow structure with a significantly lower $T_c$ on the low pressure side of the valley compared to the maximum on the high-pressure side, for all methods tested. This is in agreement with the available experimental data which show a valley structure. The location of the valley is shifted to too high pressures, by a few GPa, compared to experiment, in agreement with Ref.~\onlinecite{Wu2018}. Since the valley is located around the same pressure ($P\approx 20$ GPa) for all methods, this position is determined by the underlying DFT calculation, and apparently is reasonably well described already at this level.

It is furthermore worth to point out a second change to the $P$-$T_c$ curves. In the case of $GW$+EDMFT, the modifications are less trivial compared to the rigidly shifted $G^0W^0$, sc$GW$ and multitier $GW$+EDMFT curves, as becomes clear from Fig.~\ref{Fig:TcTheory_static}. The most notable difference is an increase in the  separation between the valley minimum and $T_c$ maximum ($\Delta T_c^{\textrm{peak}}\approx 3.5$ K) upon the inclusion of the EDMFT self-energy and polarization. This behavior is consistent with the experimental data of Guo {\it et al.}\cite{Guo2017} (the maximum, however, is located $\sim 5$ GPa too low in our calculations). In addition, at higher pressures, the $T_c$ curve starts to deviate from the monotonic decline predicted by the other methods. A more thorough discussion on the changes to the valley and high-pressure dependence observed will be presented in the next section.

One may wonder if these deviations observed only in the $GW$+EDMFT scheme are indicative of some nontrivial correlation effects, not properly captured within the current SCDFT formalism, or if they merely represent an artefact of a method which is not suitable for treating a weakly correlated system such as simple cubic phosphorus, as discussed previously. To explore this question we have extended the SCDFT formalism to also take into account correlation-induced modifications of the DFT one-particle energies by replacing them with quasi-particle energies. This approach goes beyond a treatment of correlations through the screened interaction $W$ only, and will be described in the next section.

\subsection{Quasi-particle correction}

In the simulations so far, electronic correlation effects entered through the screened interaction $W$, while the phononic contribution and the band structure were taken from the original DFT calculation. Here, we explore two more consistent schemes, without fundamentally changing the formalism. Specifically, we will update the band structure and density of states (DOS) in (i) the electronic contribution only, by replacing the DFT one-particle energies and DOS by the quasi-particle energies obtained from the solution of the quasiparticle equation, and (ii) by also approximately taking into account this change in the phononic kernel.

Since the solution of the quasi-particle equation requires knowledge of the frequency dependence of the self-energy, we use in the following calculations the $G^0W^0$ quasi-particle energies. These can be obtained without analytical continuation of $\Sigma(i\omega_n)$ to the real axis (the real-axis data are directly available from SPEX). 
 
The quasi-particle energies $\varepsilon^{QP}_{n{\bf k}}$ are calculated as the solution of the equation
\begin{equation}
\varepsilon^{QP}_{n{\bf k}} = \varepsilon^{DFT}_{n{\bf k}} + \Sigma_{n{\bf k}}(\varepsilon^{QP}_{n{\bf k}})-V^\text{XC}_{n{\bf k}}
\label{Eq:QP-equation}
\end{equation}
and define the density of states $N^{QP}(\varepsilon)$. In scheme (i), these are then substituted for $\varepsilon_{n{\bf k}}$  and $N(\varepsilon)$ in  Eqs.~\eqref{Eq:Gapeq}, \eqref{Eq:Kelstatic}, and \eqref{Eq:Keldynamic}. In scheme (ii), we additionally replace the one-particle energies in the Fermi surface integration in the calculation of $\alpha^2F$ (Eq.~(\ref{Eq:a2F})), while keeping the electron-phonon coupling constants from DFT. For a fully consistent calculation, also $g^{n{\bf k},n'{\bf k'}}_{\lambda{\bf q}}$ would need to be recalculated. This could be done within the recently developed $GW$ perturbation theory ($GW$PT),\cite{Li2019} where, in a similar manner to Eq.~\eqref{Eq:QP-equation}, the effects of the electronic self-energy correct the DFT $V^\text{XC}_{n\mathbf{k}}$ also in the calculation of the electron-phonon coupling constants $g^{n{\bf k},n'{\bf k'}}_{\lambda{\bf q}}$.
Such a treatment however goes beyond the scope of the present study.

\begin{figure}[t] 
\begin{centering}
\includegraphics[width=0.48\textwidth]{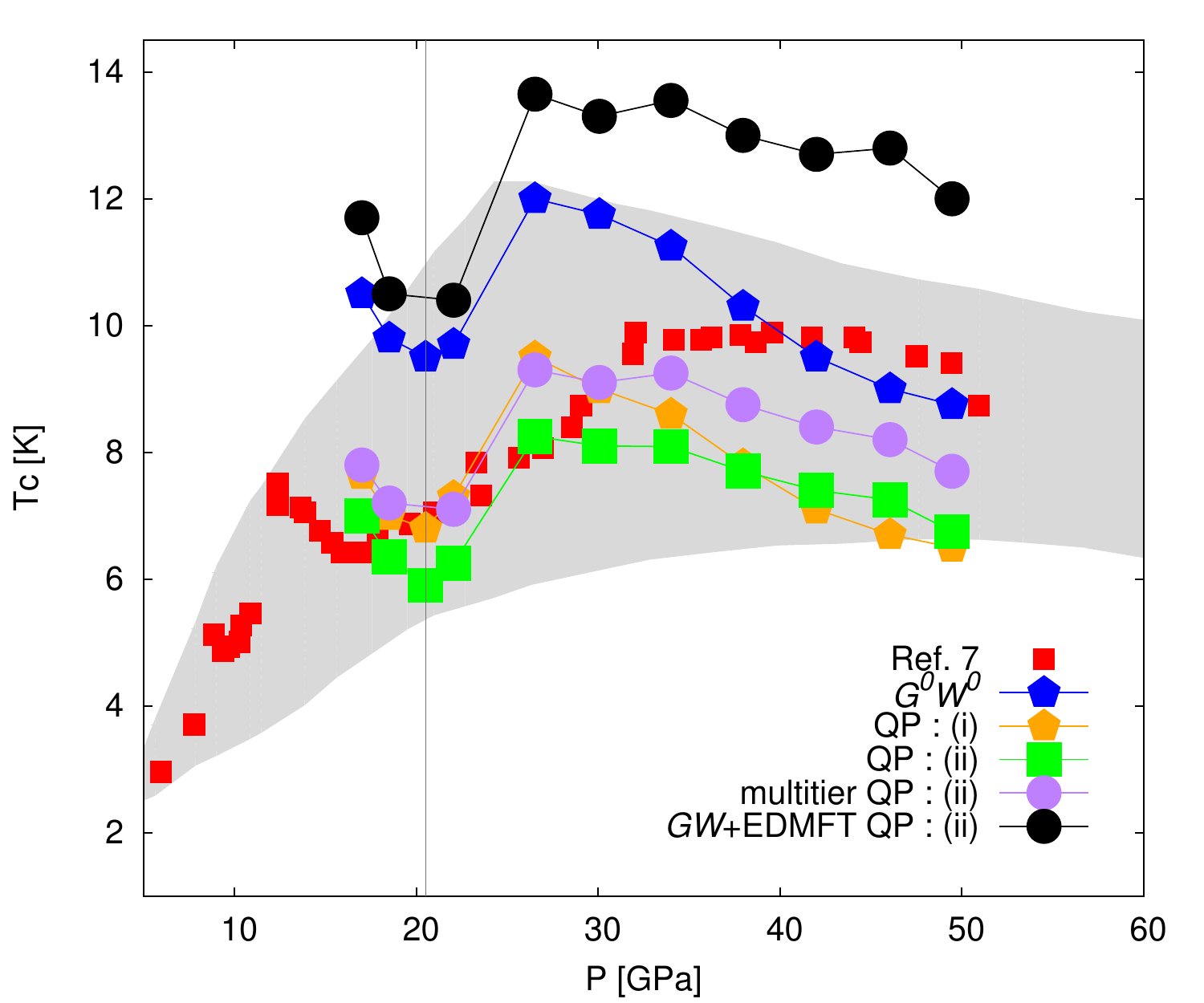}
\par\end{centering}
\caption{Correction to the pressure dependence of the theoretically calculated critical temperature $T_c$ for $G^0W^0$ from the quasi-particle bandstructure. The two methods (i) and (ii) described in the text are compared to the experimental data from Ref.~\onlinecite{Guo2017} and the original $G^0W^0$ results. Also the (multitier) $GW$+EDMFT result corrected with the $G^0W^0$ quasi-particle energies is shown for comparison. To the left of the vertical line we observe a structural instability in our phonon calculations. The full frequency dependence of the electronic contributions has been considered here. 
 \label{Fig:TcTheory_qp}}
\end{figure}

\begin{figure}[t] 
\begin{centering}
\includegraphics[width=0.48\textwidth]{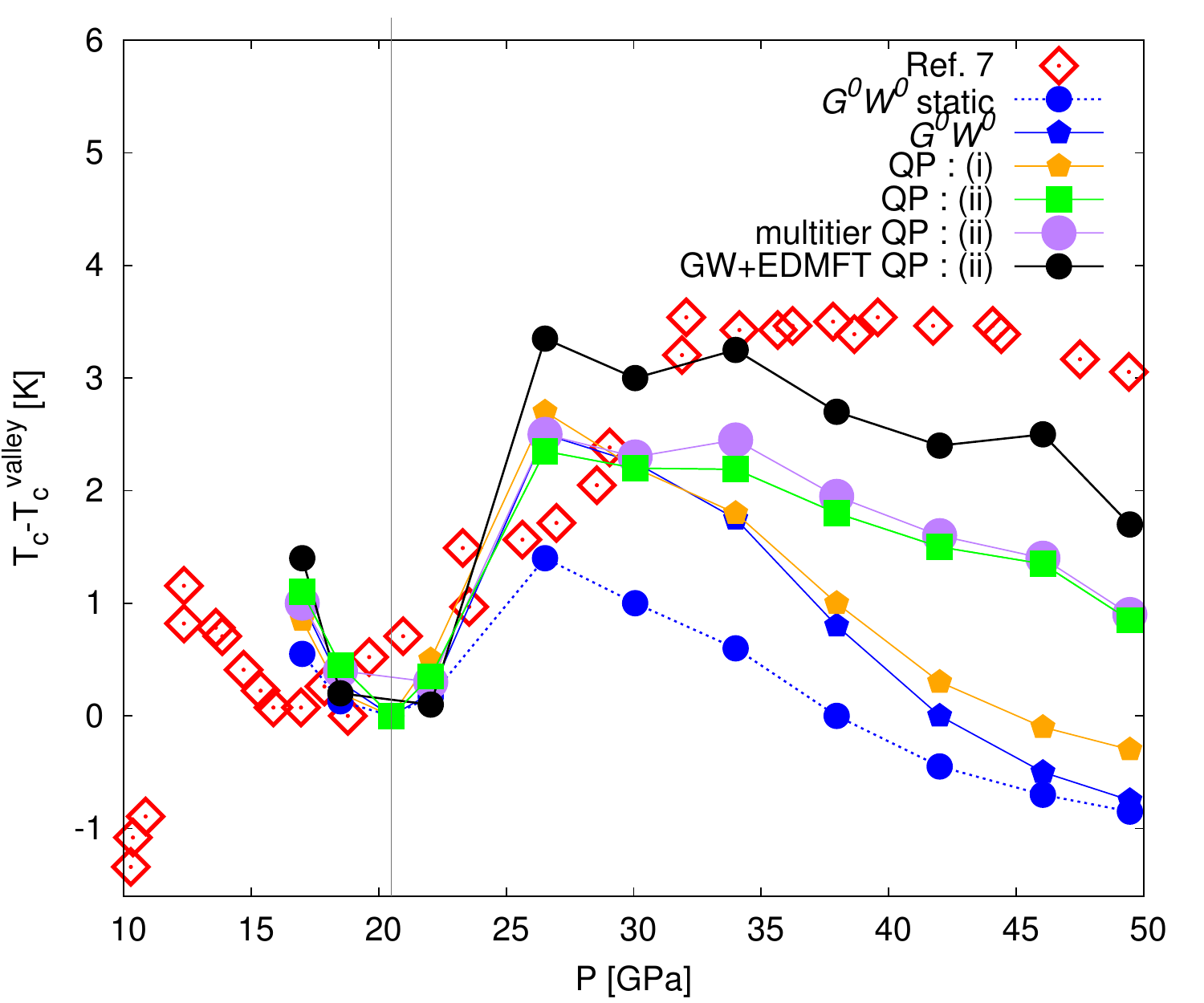}
\par\end{centering}
\caption{Calculated critical temperature $T_c$ shifted with respect to the valley minimum for the various $G^0W^0$ calculations and the (multitier) $GW$+EDMFT results corrected with the $G^0W^0$ quasi-particle energies, and comparison with the experimental data reported in Ref. \onlinecite{Guo2017}. \label{Fig:Tc relative}}
\end{figure}

Applying the procedures (i) and (ii) to one-shot $GW$, we obtain in both cases a reduction in $T_c$, so that the theoretical results are in reasonable agreement with most of the available experiments, see Fig.~\ref{Fig:TcTheory_qp} (and Fig.~\ref{Fig:ExpTc} for additional experimental results). 
In addition to a rigid shift to an overall improved $T_c$, we note that the quasi-particle correction has an additional non-trivial effect on the pressure dependence. Method (i) produces a small upwards shift in the high pressure region (orange curve), compared to one-shot $GW$ (blue curve), while the valley remains mostly unaffected. Strikingly, when also the phononic contribution is corrected in method (ii), the valley up to the peak maximum again remains approximately unaffected, while the high pressure critical temperatures above 30 GPa are significantly pushed up compared to the low-pressure region (green curve). The relative shift increases with pressure, and the resulting pressure-dependence becomes similar to the almost constant $T_c$ found in some of the experiments. For a simpler comparison of the pressure dependence, we show the $P$-$T_c$ curves shifted with respect to the valley minimum in Fig.~\ref{Fig:Tc relative}, and compare them with the experimental data from Guo {\it et al.},\cite{Guo2017} who observed a valley-ridge structure of the critical temperature.

Finally we also remark on what happens if the quasi-particle correction is applied to other methods than $G^0W^0$. For this discussion, we focus on the most advanced method considered in this work, $GW$+EDMFT.
Since $GW$+EDMFT calculations are performed on the Matsubara axis, we do not have direct access to the $\Sigma(\omega)$ needed for the solution of Eq.~\eqref{Eq:QP-equation}, and a numerical analytical continuation would be required. Here, we limit ourselves to just correcting the bandstructure and DOS in methods (i) and (ii) with the $\varepsilon^{QP}_{n{\bf k}}$ obtained from one-shot $GW$, which are known exactly.  In the SCDFT scheme, this change corresponds to using a $G^0W^0$ quasi-particle bandstructure instead of the DFT one, while ignoring further corrections to the quasi-particle energies from the EDMFT self-consistency cycle in the model space. Since the bare propagators $G^0_{\bf k}$ of the model contain a $G^0W^0$-type self-energy correction, we believe that this is more consistent than the use of the DFT bands. The incorporation of $GW$+EDMFT derived quasi-particle energies into SCDFT will be left for future work. 

Before describing the results we want to repeat that $GW$+EDMFT is arguably not the best choice for a weakly correlated materials like black phosphorus. Nevertheless, the effects of the quasi-particle correction are once again remarkable. As seen by comparing Figs.~\ref{Fig:TcTheory_qp} and \ref{Fig:TcTheory_static} (black curves), the severe overestimation of $T_c$ in the original $GW$+EDMFT is to a large extent corrected, with the theoretical predictions becoming closer to the experimental results, comparable in magnitude to the predictions from the uncorrected multitier calculations (green curve in Fig.~\ref{Fig:TcTheory_static}, note the difference in scale).
This indicates either that the quasi-particle approach somehow corrects the overestimation of $T_c$ coming from the inconsistent treatment of local and nonlocal correlations in the $s$ and $p$ subspaces, or (more likely) that the overestimation of $T_c$ is linked primarily to an inconsistency between the screened interaction $W$ and the band energies used in the SCDFT calculation. In the latter case, the lowest-order description of the nonlocal screening would then merely be responsible for the remaining modest overestimation of $T_c$. 

The most interesting effect on the $GW$+EDMFT results is, however, as in the case of $G^0W^0$, the formation of a ridge-like structure at higher pressures. While the structure of the valley remains mostly unchanged, $\Delta T_c^{\textrm{peak}}$ retains its good agreement with the experiments of Guo {\it et al.}, as demonstrated in Fig.~\ref{Fig:Tc relative}. 
The same quasi-particle correction applied to the multitier $GW$+EDMFT scheme is shown as well, with a mostly rigid shift from the $G^0W^0$ result, as in the uncorrected case. The same ridge-like high-pressure dependence is observed, and the shift to higher $T_c$ brings the theoretically calculated critical temperatures closer to the experimental reference values available in this pressure range. 

The differences between the multitier and $GW$+EDMFT results support the previous conclusion about the origin of the remaining overestimation of $T_c$ in the $GW$+EDMFT framework coming from the incorrect description of the nonlocal screening in this class of materials. 
On the other hand, the effects of the quasi-particle corrections also indicate that a better starting point than the initial DFT (GGA) calculation or more accurate quasi-particle energies are required for a quantitatively accurate description.

\section{\label{sec:Summary}Summary and Conclusions}

We have tested the reliability and predictive power of the SCDFT scheme and showed how it can be combined with the dynamically screened interactions from state-of-the-art many-body methods. By systematically studying simple cubic phosphorus under pressure, we have tested the strengths and weaknesses of different schemes for this type of weakly correlated material. Specifically, by adding the dynamic part of the electronic kernel and limiting the effects of correlations to only $W$, we do not observe any formation of a ridge in the $P$-$T_c$ dependence at high pressures, as speculated in Ref.~\onlinecite{Wu2018}. Instead a rigid shift to higher $T_c$ is found, bringing the theoretical values closer to the experimental ones. In addition to demonstrating the importance of using the full frequency dependence of the interaction to obtain meaningful estimates of the critical temperature for all the methods considered, this suggests that the peculiar pressure dependence of $T_c$ in the high-pressure region is not only of plasmonic origin. We further found that the formation of a $T_c$ valley in the simple cubic phase is predicted a few GPa too high compared to the experiments, independent of the method. This feature is therefore a result of the underlying DFT calculations and the Lifshitz transitions observed in this region, in agreement with the previous work by Wu {\it et al.}\cite{Wu2018} based on the McMillan equation.

Without corrections to the SCDFT formalism from correlations beyond $W$, the theoretical methods related to one-shot $GW$ provide the overall best agreement of $T_c$ with the available experimental data, confirming that the resummation of a subclass of diagrams in self-consistent $GW$ worsens the accuracy also for this material dependent property. Similarly, in the case of $GW$+EDMFT, the omission of higher order non-local polarization diagrams together with the more exact treatment of local contributions does not work well (specifically when applied to the $s$ and $p$ orbitals), in agreement with previous discussions related to the application of $GW$+EDMFT to weakly correlated materials.\cite{Nilsson2017} However, despite an overall too high $T_c$, the $GW$+EDMFT scheme provides the best estimate of $\Delta T_c^{\textrm{peak}}$. 

We further observed that with increasing complexity of the treatment of correlations, the pressure dependence of $T_c$ is changing from a monotonous decay (with increasing pressure) towards a valley-ridge structure, in good agreement with recent experiments predicting such a nontrivial structure.\cite{Guo2017} Modifications in the treatment of the phononic contribution have significant effects on the high pressure (25-50 GPa) $P$-$T_c$ curve, whereas the low-pressure region remains mostly unaffected. We have considered here an {\it ad-hoc} modification of the phononic kernel, which corresponds to replacing the DFT band structure with the quasi-particle energies from $G^0W^0$, and partially recalculating the phonons with these modified bands. This indicates the importance of the initial starting point for the phonon calculation. $GW$PT could give an improved description, or the use of alternative phononic kernels, such as recently proposed in Ref.~\onlinecite{Sanna2020}, may provide a viable route.

Although it is questionable if $GW$+EDMFT is a suitable method for black phosphorus, we have demonstrated a relative success in the description of the high-pressure dependence of $T_c$, especially in combination with the quasi-particle correction to the phonons.  Due to the dependence of the theoretical results on the method used to obtain the electronic kernel, we speculate that the combination of  $GW$+EDMFT and SCDFT could work well for more strongly correlated systems, where $GW$+EDMFT should provide a superior description of the fully screened electron-electron interaction, compared to the other methods considered.\cite{Boehnke2016,Nilsson2017,Petocchi2020a,Petocchi2020b,Petocchi2021} This point will be investigated in future works. To properly capture the renormalized momentum-dependent spectral function also within the SCDFT formalism, some type of quasi-particle correction would however have to be implemented in the calculation of the electronic and phononic kernels. For the phononic part, an improved starting point could be obtained from a DFT+DMFT calculation of the phonons,\cite{Savrasov2003,Kocer2020} or using DFT+$U$.\cite{Floris2011,Zhou2021}

To summarize, our results show that many-body calculations of the screened interaction $W$ in combination with the parameter-free SCDFT framework for calculating $T_c$ provides a framework which is capable of predicting the correct range of $T_c$ values in the simple cubic phase of black phosphorus. While the frequency dependence of the interaction is important for obtaining realistic $T_c$ values, it is not solely responsible for the peculiar pressure versus $T_c$ dependence that has been observed in experiments. Instead, our results indicate that it is important to use an improved phononic contribution, which goes beyond the common DFT-based kernel. To clarify whether or not quasi-particle corrections are sufficient for an accurate prediction of the $P$-$T_c$ diagram, more systematic and rigorous calculations, and additional accurate experimental reference data would be needed. 

\begin{acknowledgments}
The calculations have been performed on the Beo05 cluster at the University of Fribourg. This work was supported by ERC Consolidator Grant No. 724103 and by the Swiss National Science Foundation via NCCR Marvel and Grant No. 200021-196966.

\end{acknowledgments}

\bibliography{paper.bib}

\end{document}